\documentclass[aps,superscriptaddress,reprint]{revtex4-1}
\usepackage{graphicx}
\usepackage{dcolumn}
\usepackage{bm}
\usepackage{amsmath}
\usepackage{textcomp}
\usepackage{color}

\begin{document}

\title{Spectroscopic Investigation of Yb-doped Silica Glass for Solid-State Optical Refrigeration}

\author{Esmaeil Mobini}
\affiliation{Department of Physics \& Astronomy, University of New Mexico, Albuquerque, NM 87131, USA}
\affiliation{Center for High Technology Materials, University of New Mexico, Albuquerque, NM 87106, USA}
\author{Mostafa Peysokhan}
\affiliation{Department of Physics \& Astronomy, University of New Mexico, Albuquerque, NM 87131, USA}
\affiliation{Center for High Technology Materials, University of New Mexico, Albuquerque, NM 87106, USA}
\author{Behnam Abaie}
\affiliation{Department of Physics \& Astronomy, University of New Mexico, Albuquerque, NM 87131, USA}
\affiliation{Center for High Technology Materials, University of New Mexico, Albuquerque, NM 87106, USA}
\author{Markus P. Hehlen}
\affiliation{Department of Physics \& Astronomy, University of New Mexico, Albuquerque, NM 87131, USA}
\affiliation{Engineered Materials (MST-7), Los Alamos National Laboratory, Los Alamos, NM 87545, USA}
\author{Arash Mafi}
\email[]{mafi@unm.edu}
\affiliation{Department of Physics \& Astronomy, University of New Mexico, Albuquerque, NM 87131, USA}
\affiliation{Center for High Technology Materials, University of New Mexico, Albuquerque, NM 87106, USA}

\date{\today}

\begin{abstract}
We have argued that a high-purity Yb-doped silica glass can potentially be cooled via anti-Stokes fluorescence optical refrigeration. This conclusion
is reached by showing, using reasonable assumptions for the host material properties, that the non-radiative decay rate of Yb ions can be made substantially 
smaller than the radiative decay rate. Therefore, an internal quantum efficiency of near unity can be obtained. 
Using spectral measurements of the fluorescence emission from a Yb-doped silica optical fiber at different temperatures, we estimate 
the minimum achievable temperature in Yb-doped silica glass for different values of internal quantum efficiency.
\end{abstract}
\maketitle


\section{Introduction}
In solid-state optical refrigeration, anti-Stokes fluorescence removes thermal energy from the material, resulting in net cooling. 
Solid-state optical cooling was first proposed by Pringsheim in 1929~\cite{Pringsheim1929} and was put on a solid thermodynamic foundation 
by Landau in 1946~\cite{landau1946thermodynamics}. Solid-state optical cooling was first experimentally observed in 1995 by Epstein's 
group at Los Alamos National Laboratory in Yb-doped ZBLANP glass~\cite{epstein1995observation}. 
Much attention has since been devoted to solid-state optical refrigeration in different materials and geometries due to its interesting 
basic science properties and potential applications~\cite{epstein2010optical}. The quest for solid-state optical cooling in new configurations 
and materials is on-going~\cite{seletskiy2016laser}. 

In particular, solid-state optical refrigeration of Yb-doped silica glass, which is extensively used in high-power fiber lasers, is highly desirable. 
New generations of high power fiber amplifiers and lasers now operate at few kiloWatt levels~\cite{Richardson}. However, 
the significant heat-load in high-power operation has hindered the efforts to further scale up the power in 
fiber lasers and amplifiers~\cite{Richardson,Smith:11,Dawson:08,Jauregui:12}. 
Different methods have been developed to manage the heat-load in high-power fiber lasers or amplifiers; in particular, 
solid-state optical refrigeration via anti-Stokes fluorescence has been suggested as a viable path for heat 
mitigation~\cite{bowman1999lasers,bowman2010minimizing,Esmaeil2018josabRBL}. 
So far, there is no report of solid-state optical refrigeration
in Yb-doped silica; this manuscript is intended to highlight its possibility.
 
In this context, Radiation-Balanced Lasers (RBL) were first introduced by Bowman in 1999~\cite{bowman1999lasers}. In radiation balancing, the heat that 
originates from the quantum defect of the laser as well as parasitic absorption can be removed by anti-Stokes fluorescence under a very subtle balance condition between different 
parameters of a laser (or an amplifier)~\cite{bowman1999lasers,bowman2010minimizing,nemova2009athermal}. In other words, the anti-Stokes fluorescence removes 
the excess heat generated in the medium. Therefore, heat mitigation by radiation-balancing via anti-Stokes fluorescence is highly desirable and 
will have great practical implications if it can be achieved in Yb-doped silica glass, which is the material of choice for most high-power 
fiber lasers and amplifiers~\cite{Richardson,pask1995ytterbium,paschotta1997ytterbium}.

The investigation of solid-state optical refrigeration can be done either directly or indirectly. In a direct investigation, the material is exposed to a laser in a thermally isolated setup, often in a sophisticated vacuum environment~\cite{seletskiy2010laser}, and its temperature is measured directly by a thermal 
camera or similar methods. In an indirect method, the spectroscopic properties of materials at different temperatures are measured to evaluate the 
possibility of solid-state optical refrigeration~\cite{epstein1995observation,seletskiy2010laser,lei1998spectroscopic,melgaard2010spectroscopy}. In this manuscript, we use the indirect method to argue for the potential of high-quality Yb-doped silica glass for solid-state optical refrigeration and radiation-balancing in lasers and amplifiers.

In order to characterize the cooling potential of Yb-doped silica glass, we use the cooling efficiency $\eta_{c}$ defined as~\cite{seletskiy2010laser,melgaard2010spectroscopy}
\begin{align}
\label{Eq:cooleff}
\eta_{c}(\lambda_{p},T)=\eta_{q}\,\eta_{abs}(\lambda_{p},T) \frac{\lambda_{p}}{\lambda_{f}(T)}-1.
\end{align}
In Eq.~\ref{Eq:cooleff}, $\lambda_{f}$ is the mean fluorescence wavelength and $\lambda_{p}$ is the pump wavelength. 
$\eta_{q}$ is the internal quantum efficiency and $\eta_{abs}$ is the absorption efficiency; they are defined as
\begin{align}
\label{Eq:etaq}
&\eta_{q}=\frac{W_{r}}{W_{tot}},\quad W_{tot}=W_{r}+W_{nr},\\
\label{Eq:etaabs}
&\eta_{abs}(\lambda_{p},T)=\frac{\alpha_{r}(\lambda_{p},T)}{\alpha_{r}(\lambda_{p},T)+\alpha_{b}},
\end{align}
where $W_{r}$, $W_{nr}$, and $W_{tot}$ are radiative, non-radiative, and total decay rates of the excited state, respectively.
$\alpha_{b}$ is the background absorption coefficient, and $\alpha_{r}$ is the resonant absorption coefficient.
Note that $\alpha_{b}$ does not contain the attenuation 
due to scattering as this process does not lead to heating of the material. 
We have assumed that due to the small cross sectional area of optical fibers, the fluorescence escape efficiency to be 
unity~\cite{epstein1995observation,ruan2006enhanced}. The mean fluorescence wavelength is defined by
\begin{align}
\lambda_{f}(T)=\frac{\int_{\Delta} \lambda~S(\lambda,T) d\lambda}{\int_{\Delta} S(\lambda,T) d\lambda},
\label{Eq:meanwave}
\end{align}
where $S(\lambda,T)$ is the fluorescence power spectral density, which is a function of the glass temperature $T$, and  
$\Delta$ is the spectral domain encompassing the relevant emission spectral range~\cite{seletskiy2010laser,melgaard2010spectroscopy}. 

In order to achieve net solid-state optical refrigeration, it is necessary for the cooling efficiency to be positive. Therefore, we must show that $\eta_c>0$ is attainable over a range of $\lambda_p$ and $T$ values. It can be seen from Eq.~\ref{Eq:cooleff} that because $\lambda_p$ and $\lambda_f$ 
are often very close to each other in solid-state optical refrigeration schemes, the internal quantum efficiency $\eta_q$ has to be close to 
unity ($\eta_{q}\approx 1$)~\cite{epstein1995observation,seletskiy2010laser}. There are two main processes that lower the internal 
quantum efficiency: the multi-phonon non-radiative relaxation and the concentration quenching effect~\cite{hehlen2007model,digonnet2001rare,hoyt2003advances,barua2008influences,van1983nonradiative,auzel2003radiation,boulon2008so,nguyen2012all}. 
We will argue that the multi-phonon non-radiative relaxation is negligible in Yb-doped silica glass and the concentration quenching process can be 
prevented if the Yb ion density is kept lower than the characteristic Yb ion quenching concentration. 

In order to evaluate the absorption efficiency $\eta_{abs}$, we need to know the background absorption ($\alpha_b$) and resonant absorption ($\alpha_r$) coefficients. 
For the background absorption coefficient in Yb-doped silica glass, we will use typical values reported in the literature~\cite{jetschke2008efficient,leich2011highly,Sidharthan:18}. 
By capturing the  power spectral density of a Yb-doped silica optical fiber from its side at different temperatures, $S(\lambda,T)$, we can also obtain the resonant absorption coefficient ($\alpha_{r}$) as well as the mean fluorescence wavelength ($\lambda_{f}$). Therefore, we will have all the necessary parameters in 
Eq.~\ref{Eq:cooleff}; using this information we will estimate the cooling efficiency and show that solid-state optical refrigeration is feasible 
in Yb-doped silica glass.
\section{Internal quantum efficiency}
The internal quantum efficiency is the fraction of the radiative decay versus the total decay of an excited state in a medium;
therefore, the presence of non-radiative decay channels characterized by the non-radiative decay rate $W_{\rm nr}$ in Eq.~\ref{Eq:etaq}
are responsible for decreasing $\eta_q$ below unity. The non-radiative decay channels in a typical Yb-doped silica glass can be broken 
down according to the following equation: 
\begin{align}
W_{\rm nr}=W_{\rm mp}+W_{\rm OH^{-}}+W_{\rm Yb}+\sum_{\rm TM} W_{\rm TM}+\sum_{\rm RE} W_{\rm RE}.
\label{Eq:total-nonradiative}
\end{align}
The partial non-radiative decay channels are as follows: 
$W_{\rm mp}$ represents the multi-phonon decay of the Yb excited state, 
$W_{\rm OH^{-}}$ accounts for non-radiative decay of the Yb excited state via the high-energy vibrational modes of ${\rm OH^{-}}$ impurities, 
$W_{\rm Yb}$ accounts for non-radiative decay in Yb ion clusters, 
and $W_{\rm TM}$ and $W_{\rm RE}$ represent non-radiative decay due to interactions of the excited state with various 
transition-metal and rare-earth ion impurities, respectively.

\begin{figure}[!h]
    \includegraphics[width=3.4 in]{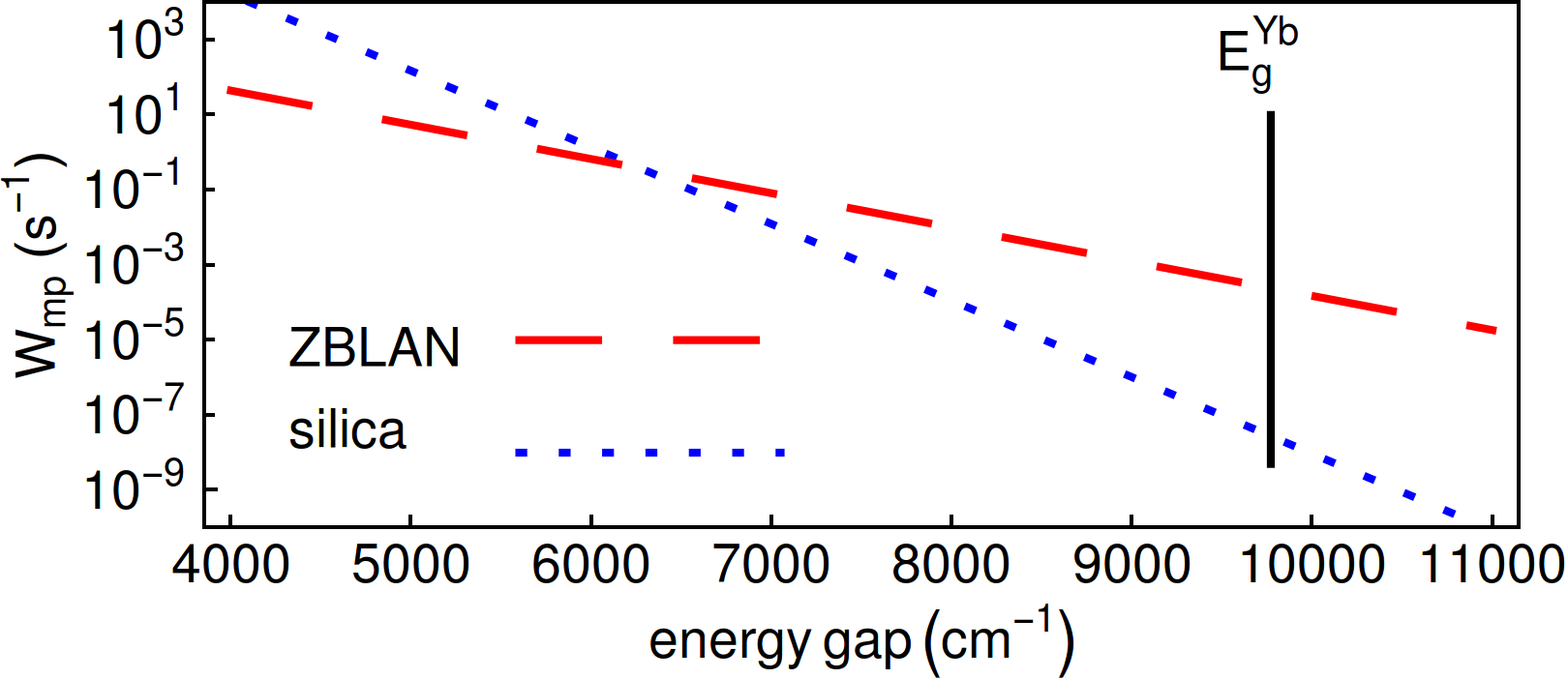}
\caption{Multi-phonon non-radiative decay rate ($W_{mp}$) of Yb-doped ZBLAN and silica glasses versus energy gap ($E_{g}$)
calculated from Eq.~\ref{Eq:multi-phonon decay} and the parameters listed in Table~\ref{silicazblan}.}
\label{Fig:omeganr}
\end{figure}

In the following, we will discuss the various non-radiative decay channels in Eq.~\ref{Eq:total-nonradiative} and show that 
they can be made sufficiently small to allow for a near-unity internal quantum efficiency value ($\eta_q\approx 1$). 
We first begin with the multi-phonon relaxation that originates from the coupling of the excited state with the vibrational 
wavefunctions of the ground state. 
Using the energy-gap law~\cite{van1983nonradiative,faure2007improvement,hehlen2007model,hoyt2003advances}, we can calculate the decay rate from 
\begin{align}
\label{Eq:multi-phonon decay}
W_{\rm mp}=W_{0}~e^{-\alpha_{h} ( E_{g}-2E_{p})},
\end{align}
where $E_{p}$ is the maximum phonon energy of the host material, and $E_{g}$ is the energy gap of the dopant ion (Yb).
$W_{0}$ and $\alpha_{h}$ are phenomenological parameters, whose values strongly depend on 
 the host-material~\cite{van1983nonradiative,faure2007improvement,hehlen2007model,hoyt2003advances}. 
Figure~\ref{Fig:omeganr} shows the multi-phonon non-radiative decay rates of silica and ZBLAN glasses versus the 
energy gaps of the doped ions at $T\,=\,300$\,K, using the parameters shown in Table~\ref{silicazblan}.
\begin{table}[htp]
   \caption{Parameters related to Eq.~\ref{Eq:multi-phonon decay} and Fig.~\ref{Fig:omeganr} for silica and ZBLAN glasses~\cite{faure2007improvement,hehlen2007model,hoyt2003advances}.}
\begin{center}
\label{silicazblan}
 \renewcommand{\arraystretch}{1.3}
\begin{tabular}{ | p{13mm} | p{17mm} | p{17mm} | p{17mm} |}
      \hline
      Host & $W_{0}\,({\rm s}^{-1})$ & $\alpha_h\,({\rm cm})$ & $E_p\,({\rm cm}^{-1})$ \\ 
      \hline
       silica & $7.8 \times 10^{7}$   & $4.7 \times 10^{-3}$ & $1.10 \times 10^{3}$ \\
      \hline
      {\small ZBLAN} & $1.7 \times 10^{4}$ & $2.1 \times 10^{-3}$ & $0.58 \times 10^{3}$ \\ \hline
\end{tabular}
\end{center}
\end{table}

The vertical solid line in Fig.~\ref{Fig:omeganr} marks the energy gap of a ${\rm Yb^{3+}}$ ion.
It is evident that for Yb-doped silica glass, the non-radiative decay rate is around $W_{\rm mp}^{\rm silica} \approx 10^{-8}~s^{-1}$, which is much smaller than 
the Yb-doped ZBLAN glass multi-phonon decay rate $W_{\rm mp}^{\rm ZBLAN} \approx 10^{-4}~s^{-1}$. This comparison suggests that with 
respect to Yb multi-phonon relaxation, silica glass is a more suitable choice for solid-state optical refrigeration than ZBLAN glass.

Considering the advances in materials synthesis of fiber preforms, the term $W_{\rm OH^{-}}$ in Eq.~\ref{Eq:total-nonradiative} can be made 
very small (see e.g. dry fiber technology~\cite{thomas2000towards}); therefore, it can be neglected~\cite{barua2008influences}. It has also been 
shown by Auzel et al.~\cite{auzel2003radiation} that the total effect of the last three terms in Eq.~\ref{Eq:total-nonradiative}, 
$W_{\rm Yb}+\Sigma W_{\rm TM}+\Sigma W_{\rm RE}$, can be described by a phenomenological equation based on a limited diffusion process, 
modeled as a non-radiative dipole-dipole interaction between the ions and impurities~\cite{auzel2003radiation,boulon2008so}. 
This concentration quenching process can be prevented if the Yb ion density is lower than the critical quenching concentration of the Yb-doped silica glass, 
which exists because there are impurities. Therefore, the critical quenching concentration is generally a sample specific quantity. That is, it would be higher 
for lower impurity concentrations.
For a Yb ion density smaller than the critical quenching concentration, the internal quantum efficiency can approach 
$\eta_{q}\approx 1$~\cite{auzel2003radiation,boulon2008so,nguyen2012all}. 
It must be noted that an internal quantum efficiency of $\eta_{q}$\,=\,0.95 is reported in \cite{jeong2004ytterbium} for Yb-doped silica, which is 
consistent with our claim that $W_{\rm nr}$ can be made quite small in Yb-doped silica.
\section{Absorption efficiency and mean fluorescence wavelength}
\begin{figure}[!h]
    \includegraphics[width=3.4 in]{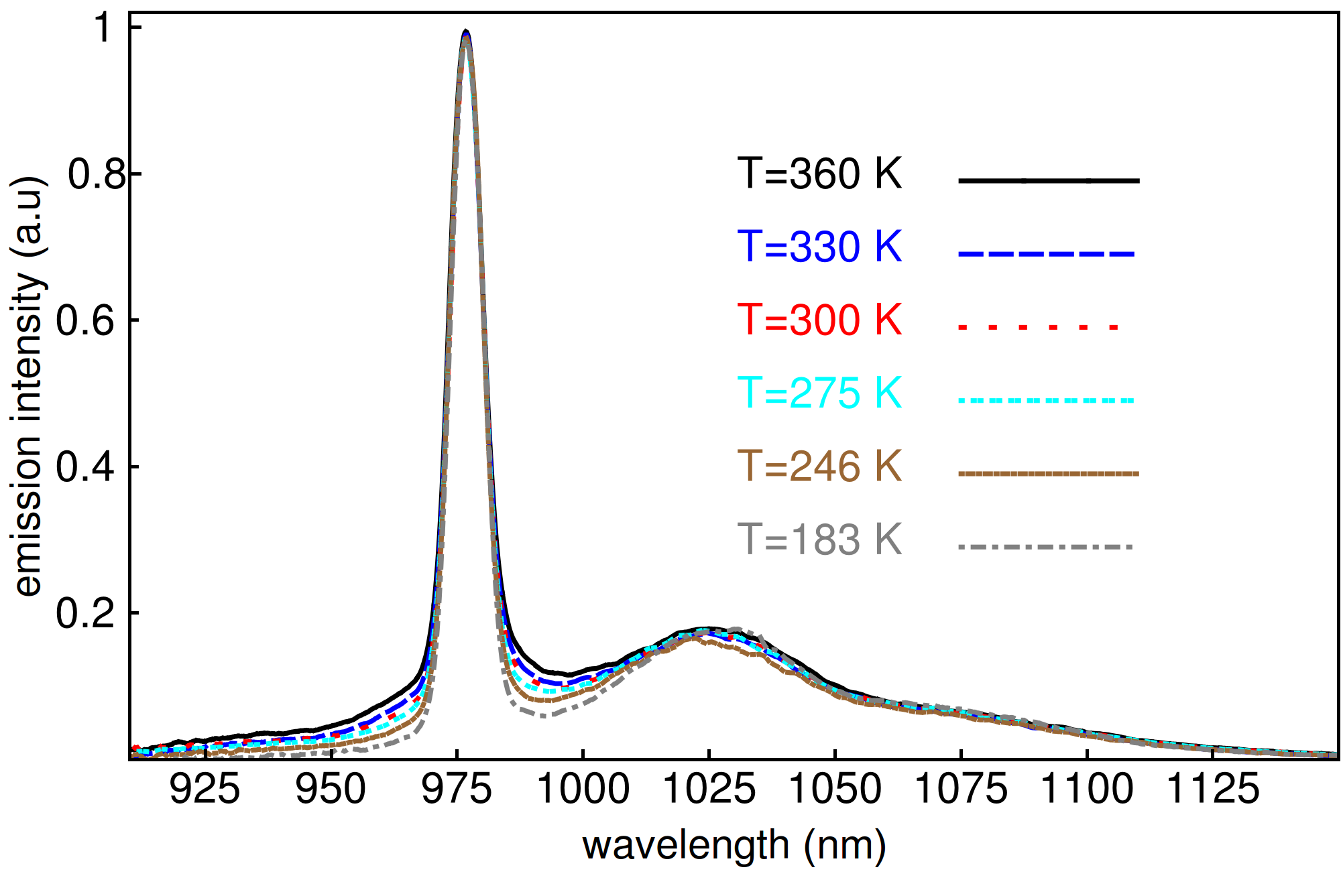}
\caption{Measured peak normalized emission spectra of DF-1100 Yb-doped silica fiber at different temperatures.}
\label{Fig:NormEmission}
\end{figure}

In order to calculate the cooling efficiency, we still need to obtain the resonant absorption coefficient and the mean 
fluorescence wavelength, both of which can be obtained from a spectroscopic investigation. The resonant absorption coefficient
is used in conjunction with Eq.~\ref{Eq:etaabs} to determine the absorption efficiency. The setup implemented in our 
experiment consists of a single-mode Yb-doped silica fiber (DF-1100, from Newport Corporation) that is pumped by a 
Ti:Sapphire laser at $\lambda$\,=\,900\,nm. The fiber is mounted on a plate whose temperature is changed from nearly 180\,K up to 360\,K. 
The fluorescence of the Yb-doped silica fiber is captured by a multimode fiber from the side of the Yb-doped silica fiber and 
is sent to an Optical Spectrum Analyzer. Figure.~\ref{Fig:NormEmission} shows the measured 
fluorescence spectra (power spectral density $S(\lambda,T)$), normalized to their peak values at $\lambda_{\rm peak}\,\approx$\,976\,nm, at different temperatures. 

By inserting the measured fluorescence spectra into Eq.~\ref{Eq:meanwave} and considering $\Delta\in\{905{\rm nm},1150{\rm nm}\}$, 
the dependence of the mean fluorescence wavelength on temperature is obtained. The mean fluorescence wavelength follows 
approximately the following function:
\begin{align}
\label{Eq:lambdafT}
\lambda_{f}(T) \approx 999~({\rm nm})\,+\,b\times T^{-1}, \quad b\,=\,2735\pm\, 31 {\rm nm/K}.
\end{align}
This behavior at temperatures above 245\,K to 360\,K is 
nearly linear, which is similar to that reported in other host materials, such as ZBLAN~\cite{hehlen2007model,lei1998spectroscopic}.

In order to calculate the the resonant absorption coefficient $\alpha_{r}$, we first calculate the emission cross section $\sigma_{e}$, 
and then use the McCumber relation to obtain the absorption cross section $\sigma_{a}$ and then the resonant absorption 
coefficient $\alpha_{r}$~\cite{fan1989end,mccumber1964theory,newell2007temperature}. The emission cross section is obtained from the 
measured fluorescence power spectral density $S(\lambda,T)$ via the F\"uchtbauer-Ladenburg equation~\cite{aull1982vibronic,newell2007temperature}:
\begin{align}
\sigma_{e}(\lambda,T)=\dfrac{\lambda^5}{8\,\pi\,n^2\,c\,\tau_{r}(T)}\times \dfrac{S(\lambda,T)}{\int_{\Delta} \lambda~S(\lambda,T) d\lambda },
\label{Eq:FuchtbauerLadenburg}
\end{align}
where $n$ is the refractive index of the fiber core, $c$ is the speed of light in vacuum, and $\tau_{r}=W_{r}^{-1}$ is the radiative lifetime.

\begin{figure}[!h]
    \includegraphics[width=3.4 in]{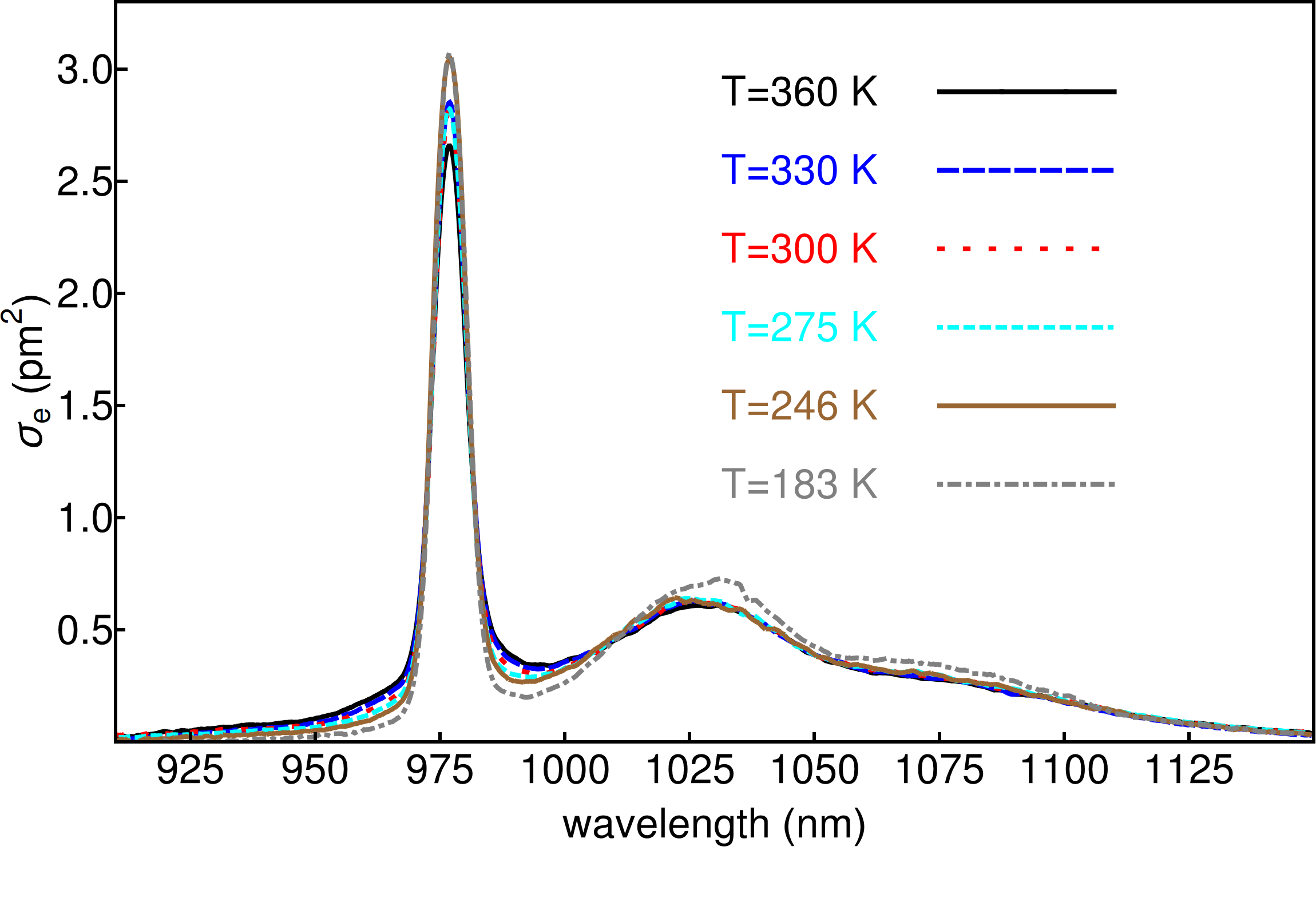}
\caption{Emission cross section versus wavelength for DF-1000 Yb-doped silica fiber at different temperatures. 
The spectra were calculated from Eq.~\ref{Eq:FuchtbauerLadenburg} using the emission spectra shown in Fig.~\ref{Fig:NormEmission} 
and the measured radiative lifetimes from Ref.~\cite{Mobini:18}.}
\label{Fig:emission-cross}
\end{figure}

\begin{figure}[!h]
    \includegraphics[width = 3.4 in]{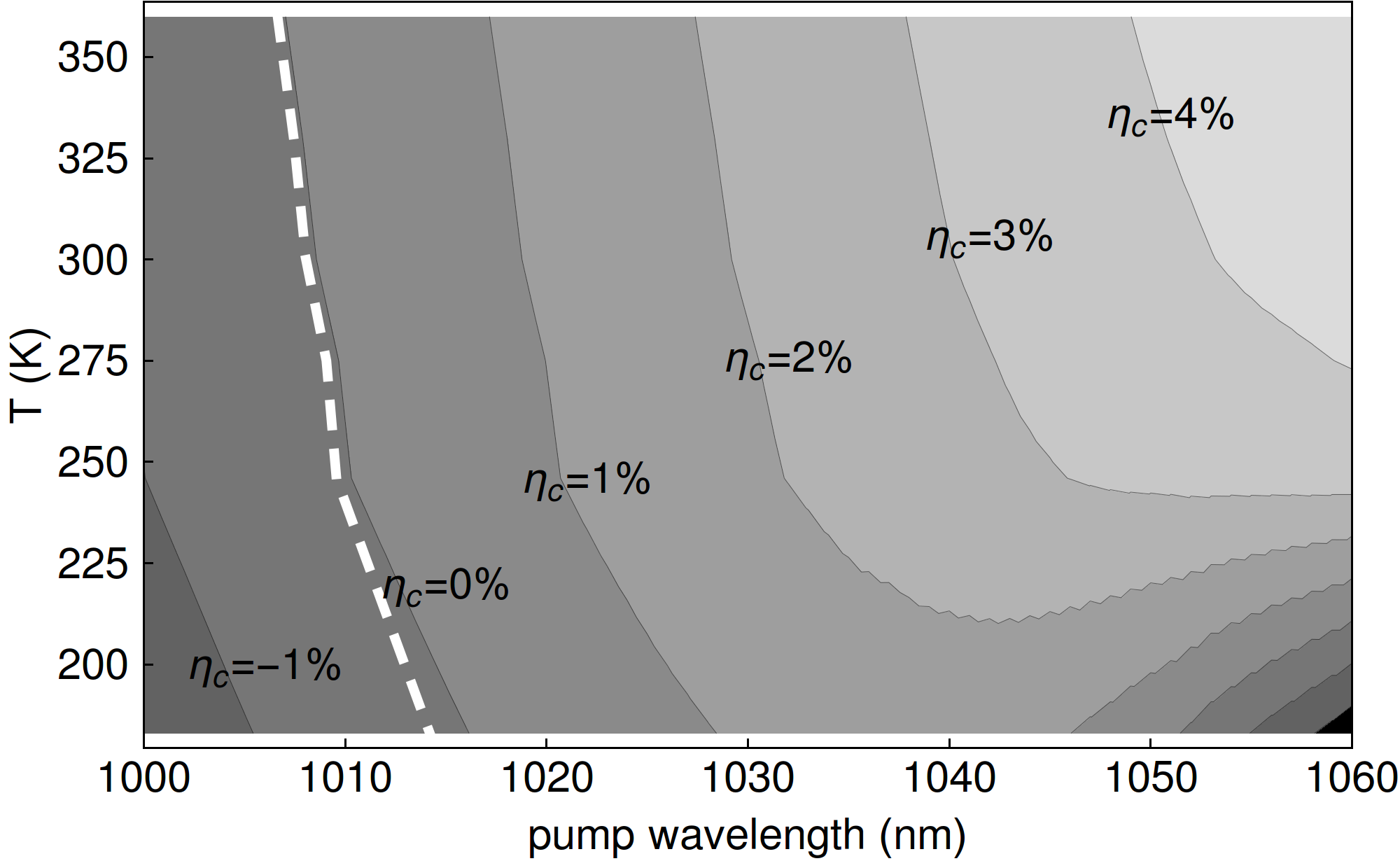}
\caption{Cooling efficiency versus temperature and pump wavelength with $\eta_{q}$\,=\,1 and $\alpha_{b}\,=\,10$\,dB/km
for DF-1100 Yb-doped silica fiber calculated from Eq.~\ref{Eq:cooleff}. The dashed line connects the experimental measurements of the mean 
fluorescence wavelength versus the temperature.}
\label{Fig:2PD}
\end{figure}

In order to apply Eq.~\ref{Eq:FuchtbauerLadenburg}, the radiative lifetime at each temperature needs to be measured. In high-quality samples
for which the non-radiative decay rates are negligible compared to the radiative decay rates, the fluorescence lifetimes are comparable 
to the radiative lifetimes ($\tau_{f}\approx \tau_{r}$); therefore, we measured the fluorescence lifetimes at different temperatures from the 
side of the fiber~\cite{Mobini:18}. Using this assumption, the emission cross sections at different temperatures were calculated and
are shown in Fig.~\ref{Fig:emission-cross}. The absorption cross sections can be readily obtained using the McCumber relation:
\begin{align}
\label{Eq:McCumber}
&\sigma_{a}(\lambda,T)=\sigma_{e}(\lambda,T)\times \mathcal{Z}(\lambda,T),\\
\nonumber
&\mathcal{Z}(\lambda,T)\,=\,\exp\left[\frac{hc}{k_{b}T}(\frac{1}{\lambda}-\frac{1}{\lambda_{0}})\right],
\end{align}
where $k_{b}$ is the Boltzmann constant, $h$ is the Planck constant and $\lambda_{0}\,=\,976$\,nm is the wavelength corresponding 
to the zero-line phonon energy~\cite{mccumber1964theory,melgaard2010spectroscopy,pask1995ytterbium}.
The resonant absorption coefficient can be calculated from $\sigma_{a}(\lambda,T)$ in Eq.~\ref{Eq:McCumber} (and Fig.~\ref{Fig:emission-cross}) using 
\begin{align}
\label{Eq:alphardef}
\alpha_{r}(\lambda,T)\,=\,\sigma_{a}(\lambda,T)\times N.
\end{align}
Here, we will assume a typical Yb ion density of $N\,=5\times 10^{25}\,{\rm m}^{-3}$.
We now have all the necessary ingredients to calculate the cooling efficiency $\eta_{c}$ in Eq.~\ref{Eq:cooleff}. We only need to
provide a value for the background absorption coefficient in Eq.~\ref{Eq:etaabs} to determine the absorption efficiency $\eta_{abs}$.
Here, we assume a background absorption coefficient of $\alpha_{b}\,=\,10$\,dB/km\,$\approx 2.3\times~10^{-3}$/m, which is a typical value for commercial grade 
Yb-doped silica fibers. Using this information,  we present a contour plot of the cooling efficiency $\eta_{c}$ in Fig.~\ref{Fig:2PD}
as a function of the pump wavelength and temperature, assuming that $\eta_{q}$\,=\,1. Note that we only know the values of $\alpha_{r}(\lambda,T)$ 
at discrete values of temperature $T$ for which our measurements were performed in Fig.~\ref{Fig:NormEmission}; the density plot in 
Fig.~\ref{Fig:2PD} is an interpolation of the measured values.
It is seen in Fig.~\ref{Fig:2PD} that with a decrease in the temperature, the cooling efficiency decreases; 
this behavior is due to the red-shift of the mean fluorescence wavelength and the decrease in the resonant absorption 
coefficient with decreasing temperature~\cite{seletskiy2010laser,hehlen2007model}. 

\begin{figure}[!h]
    \includegraphics[width=3.4 in]{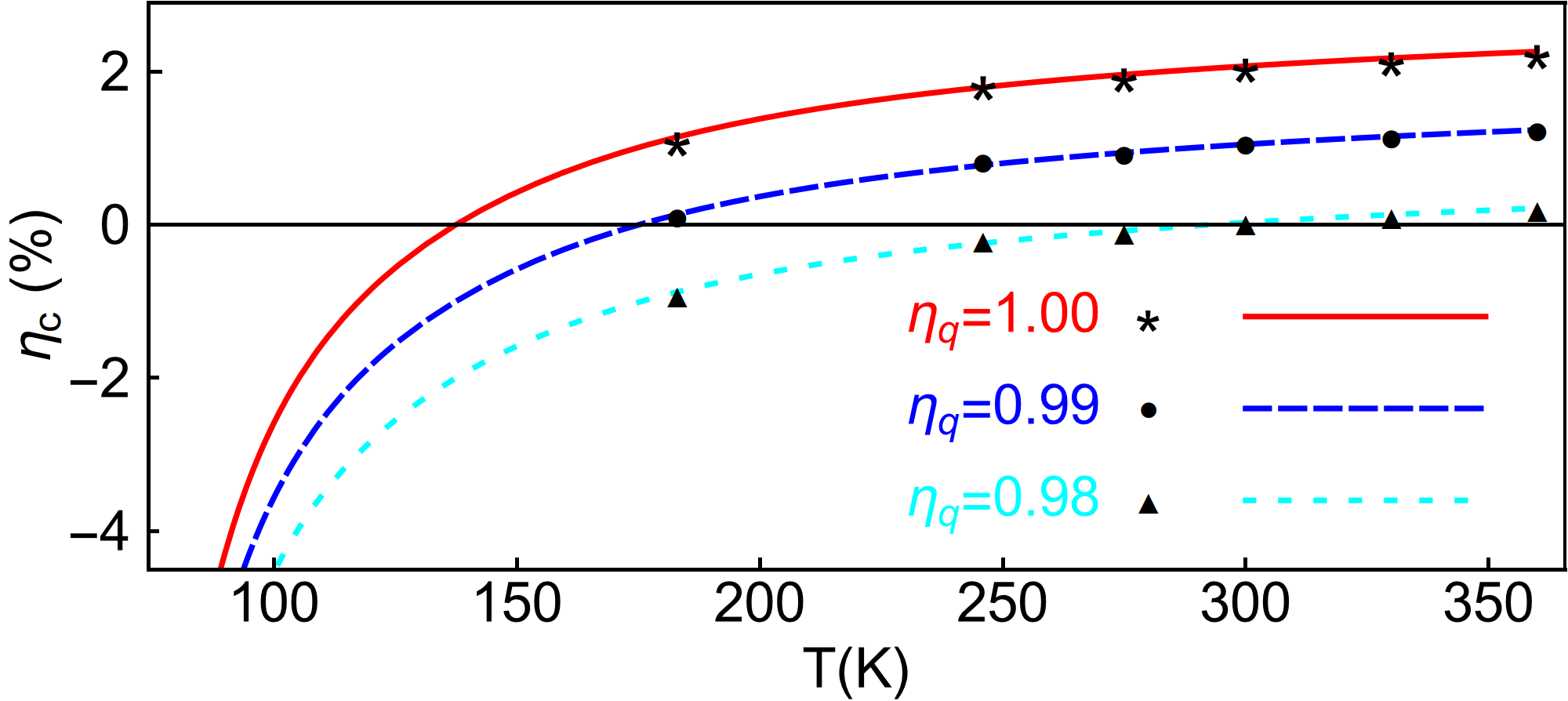}
\caption{Cooling efficiency for different values of quantum efficiency versus temperature with $\alpha_{b}\,=\,10$\,dB/km
for DF-1100 Yb-doped silica fiber calculated from Eq.~\ref{Eq:cooleff} for different internal quantum efficiencies, $\eta_q$. 
The colored lines are plotted using Eq.~\ref{Eq:cooleff} and the fitting presented in Eq.~\ref{Eq:Fit3}.}
\label{Fig:cool-eff}
\end{figure}

In practice, it is impossible to achieve an internal quantum efficiency of unity; therefore, in Fig.~\ref{Fig:cool-eff} we investigate 
the effect of a non-ideal internal quantum efficiency on the cooling efficiency, for $\lambda_{p}\,=\,1030$\,nm, as a function of the temperature.
The discrete points in Fig.~\ref{Fig:cool-eff} signify the values of $\eta_{c}$ obtained for the assumed $\eta_{q}$ at the particular measured temperatures
reported in Fig.~\ref{Fig:NormEmission}. The apparent difference between the cooling efficiency obtained for $\eta_{q}$\,=\,1 versus $\eta_{q}$\,=\,0.98 
highlights the importance of having a high-quality glass for radiative cooling. While the discrete points in Fig.~\ref{Fig:cool-eff} reveal the
main expected behavior of $\eta_{c}$ versus the temperature, it is helpful to estimate the minimum achievable temperature for solid-state
 optical refrigeration in Yb-doped silica glass, subject to the assumptions made about $\eta_q$, $N$, and $\alpha_b$. In order to do so, we next present an analytical fitting to the discrete points in Fig.~\ref{Fig:cool-eff} that can be used to estimate the minimum achievable temperature. 
The analytical fitting, which is described in the next paragraph, is used in conjunction with Eq.~\ref{Eq:cooleff} 
to plot the colored lines for each value of $\eta_q$ in Fig.~\ref{Fig:cool-eff} and is in reasonable agreement with 
the experimentally measured discrete data.  
From the discussions above and Eqs.~\ref{Eq:FuchtbauerLadenburg},~\ref{Eq:McCumber}, and~\ref{Eq:alphardef}, 
we note that $\alpha_{r}(\lambda_p,T)$ (at the pump wavelength) can be expressed as:
\begin{align}
\alpha_{r}(\lambda_p,T) \propto \frac{1}{c} \frac{\lambda_p^5}{\tau_r(T)} \times \frac{S(\lambda_p,T)}{\int_{\Delta} \lambda~S(\lambda,T) d\lambda}\times \mathcal{Z}(\lambda_p,T).
\label{Eq:Fit1}
\end{align}
In Ref.~\cite{Mobini:18}, we performed fluorescence lifetime measurements in Yb-silica. Here, we present a fitting of 
$\tau_r(T)$ to an analytical form that is based on a two-level excited state:
\begin{align}
\tau_r(T)=\frac{1+\exp(-\delta E/k_b T)}{\tau^{-1}_1+\tau^{-1}_2\exp(-\delta E/k_b T)}.
\label{Eq:radiative-func}
\end{align}
$\tau_1\,=\,798\pm 2$\,\textmu s, and $\tau_2\,=\,576\pm 27$\,\textmu s are the lifetimes of the first and second energy levels of the 
excited state, respectively, and $\delta E\,=\,506\pm 56\,{\rm cm^{-1}}$ is the energy difference between these two levels~\cite{zhang1993thermal,newell2007temperature}.
We also present the following approximation:
\begin{align}
 \frac{\lambda_p^2\ S(\lambda_p,T)}{\int_{\Delta} \lambda~S(\lambda,T) d\lambda} \approx 7.4\,+\,\left(\frac{d}{T}\right)^5,
\quad d\,=\,205.9 \pm 2.4\,{\rm K}.
\label{Eq:Fit2}
\end{align}
Using Eqs.~\ref{Eq:radiative-func} and~\ref{Eq:Fit2}, we can approximate $\alpha_{r}(\lambda_p,T)$ [Eq.~\ref{Eq:Fit1}] with the following mathematical form:
\begin{align}
\alpha_{r}(\lambda_p,T) \approx \frac{\alpha_{r,0}}{c~\tau_r(T)} \times \Big(7.4\,+\,(d/T)^5\Big) \times \mathcal{Z}(\lambda_p,T).
\label{Eq:Fit3}
\end{align}

Fitting the analytical in Eq.~\ref{Eq:Fit3} to the discrete points in Fig.~\ref{Fig:cool-eff}, we find the dimensionless coefficient 
$\alpha_{r,0}=\,(0.95\pm\,0.01)\times 10^{6}$. The fitted lines in Fig.~\ref{Fig:cool-eff} show that the minimum achievable temperature can reach down to 
$T_{\rm min}\,=\,138$\,K for $\eta_q\,=\,1$, $T_{\rm min}\,=\,175$\,K for $\eta_q\,=\,0.99$, and $T_{\rm min}\,=\,290$\,K for $\eta_q\,=\,0.98$. 
Figure~\ref{Fig:cool-eff} also shows that the maximum cooling efficiency for Yb-silica glass is around $\eta^{\rm max}_{c}\approx$2\% at room 
temperature for $\lambda_p$\,=\,1030\,nm. 
Setting the background absorption to zero ($\alpha_b$\,=\,0) increases this value to $\eta^{\rm max}_{c}\approx$2.2\%. In order to increase the cooling efficiency, 
the background absorption must be minimized, the internal quantum efficiency has to be close to unity, and the ion dopant density $N$ must be increased to enhance the 
resonant absorption coefficient. We note that these requirements are not necessarily compatible with each other; e.g. increasing $N$ can potentially decrease $\eta_q$
due to quenching. Therefore, a compromise determined by careful measurements must be obtained.
\section{Discussion and Conclusion}
It must be noted that by taking $N\,=5\times 10^{25}\,{\rm m}^{-3}$ in this manuscript, we have implicitly assumed that the silica glass host is co-doped with some 
modifiers like ${\rm Al_2O_3}$, in order to shift the quenching concentration to higher values to reduce clustering and ensure an adequate cooling efficiency~\cite{arai1986aluminum,laegsgaard2002dissolution}.
For pure silica, applying the model developed by Auzel et al.~\cite{auzel2003radiation} to the experimental data from Ref.~\cite{barua2008influences}, it can be shown that
$N\,=0.7\times 10^{25}\,{\rm m}^{-3}$ can guarantee a near unity internal quantum efficiency~\cite{arai1986aluminum,laegsgaard2002dissolution}. 
Using $N=0.7 \times 10^{25} m^{-3}$ in pure silica, we have calculated the minimum achievable temperature to be
$T_{\rm min}\,=\,216$\,K for $\eta_q\,=\,1$, and $T_{\rm min}\,=\,262$\,K for $\eta_q\,=\,0.99$. For $\eta_q\,=\,0.98$, $T_{\rm min}$ is higher
than the room temperature. As expected, a decrease in ion density results in a lower cooling efficiency.

In conclusion, we have argued that a high-purity Yb-doped silica glass can potentially be cooled via anti-Stokes fluorescence optical refrigeration.
We show that, in principle, the non-radiative decay rate $W_{nr}$ can be made substantially smaller than the radiative decay rate $W_r$. 
Therefore, an internal quantum efficiency of near unity can be obtained, making 
Yb-doped silica glass suitable for solid-state optical refrigeration. Our assessment is based on reasonable assumptions for material properties, e.g. 
we have assumed a typical background absorption coefficient of $\alpha_{b}$\,=\,10\,dB/km and an
internal quantum efficiency of larger than $\eta_{q}$\,=\,0.98. We have made spectral measurements of the fluorescence from a Yb-doped silica 
optical fiber at different temperatures. Using these measurements, we have reported the temperature dependence of the mean fluorescence wavelength, 
and have estimated the minimum achievable temperature in Yb-doped silica glass. Our analysis highlights the potential for Yb-doped silica glass to be used as the
gain medium for radiation-balanced high-power fiber lasers and amplifiers.
\section*{Acknowledgment}
This material is based upon work supported by the Air Force Office of Scientific Research under award number FA9550-16-1-0362
titled Multidisciplinary Approaches to Radiation Balanced Lasers (MARBLE).


\begin{thebibliography}{10}
\newcommand{\enquote}[1]{``#1''}

\bibitem{Pringsheim1929}
P.~Pringsheim, \enquote{Zwei bemerkungen {\"u}ber den unterschied von
  lumineszenz- und temperaturstrahlung,} {{Zeitschrift
  f{\"u}r Physik}} \textbf{57}, 739--746 (1929).

\bibitem{landau1946thermodynamics}
L.~Landau, \enquote{On the thermodynamics of photoluminescence,}
  {{J. Phys. (Moscow)}} \textbf{10} (1946).

\bibitem{epstein1995observation}
R.~I. Epstein, M.~I. Buchwald, B.~C. Edwards, T.~R. Gosnell, and C.~E. Mungan,
  \enquote{Observation of laser-induced fluorescent cooling of a solid,}
  {{Nature}} \textbf{377}, 500 (1995).

\bibitem{epstein2010optical}
R.~I. Epstein and M.~Sheik-Bahae, \emph{Optical refrigeration: science and
  applications of laser cooling of solids} (John Wiley \& Sons, 2010).

\bibitem{seletskiy2016laser}
D.~V. Seletskiy, R.~Epstein, and M.~Sheik-Bahae, \enquote{Laser cooling in
  solids: advances and prospects,} {{Reports on Progress
  in Physics}} \textbf{79}, 096401 (2016).

\bibitem{Richardson}
D.~J. Richardson, J.~Nilsson, and W.~A. Clarkson, \enquote{High power fiber
  lasers: current status and future perspectives,} {{J.
  Opt. Soc. Am. B}} \textbf{27}, B63--B92 (2010).

\bibitem{Smith:11}
A.~V. Smith and J.~J. Smith, \enquote{Mode instability in high power fiber
  amplifiers,} {{Opt. Express}} \textbf{19}, 10180--10192
  (2011).

\bibitem{Dawson:08}
J.~W. Dawson, M.~J. Messerly, R.~J. Beach, M.~Y. Shverdin, E.~A. Stappaerts,
  A.~K. Sridharan, P.~H. Pax, J.~E. Heebner, C.~W. Siders, and C.~Barty,
  \enquote{Analysis of the scalability of diffraction-limited fiber lasers and
  amplifiers to high average power,} {{Opt. Express}}
  \textbf{16}, 13240--13266 (2008).

\bibitem{Jauregui:12}
C.~Jauregui, T.~Eidam, H.-J. Otto, F.~Stutzki, F.~Jansen, J.~Limpert, and
  A.~T\"{u}nnermann, \enquote{Physical origin of mode instabilities in
  high-power fiber laser systems,} {{Opt. Express}}
  \textbf{20}, 12912--12925 (2012).

\bibitem{bowman1999lasers}
S.~R. Bowman, \enquote{Lasers without internal heat generation,}
  {{IEEE journal of quantum electronics}} \textbf{35},
  115--122 (1999).

\bibitem{bowman2010minimizing}
S.~R. Bowman, S.~P. O'Connor, S.~Biswal, N.~J. Condon, and A.~Rosenberg,
  \enquote{Minimizing heat generation in solid-state lasers,}
  {{IEEE Journal of Quantum Electronics}} \textbf{46},
  1076--1085 (2010).

\bibitem{Esmaeil2018josabRBL}
E.~Mobini, M.~Peysokhan, B.~Abaie, and A.~Mafi, \enquote{Thermal modeling, heat
  mitigation, and radiative cooling for double-clad fiber amplifiers,}
  {{JOSA B}} \textbf{Early Posting}, 0000--0000 (2018).

\bibitem{nemova2009athermal}
G.~Nemova and R.~Kashyap, \enquote{Athermal continuous-wave fiber amplifier,}
  {{Optics Communications}} \textbf{282}, 2571--2575
  (2009).

\bibitem{pask1995ytterbium}
H.~Pask, R.~J. Carman, D.~C. Hanna, A.~C. Tropper, C.~J. Mackechnie, P.~R.
  Barber, and J.~M. Dawes, \enquote{Ytterbium-doped silica fiber lasers:
  versatile sources for the 1-1.2~ $\mu m$ region,} {{IEEE
  Journal of Selected Topics in Quantum Electronics}} \textbf{1}, 2--13 (1995).

\bibitem{paschotta1997ytterbium}
R.~Paschotta, J.~Nilsson, A.~C. Tropper, and D.~C. Hanna,
  \enquote{Ytterbium-doped fiber amplifiers,} {{IEEE
  Journal of quantum electronics}} \textbf{33}, 1049--1056 (1997).

\bibitem{seletskiy2010laser}
D.~V. Seletskiy, S.~D. Melgaard, S.~Bigotta, A.~Di~Lieto, M.~Tonelli, and
  M.~Sheik-Bahae, \enquote{Laser cooling of solids to cryogenic temperatures,}
  {{Nature Photonics}} \textbf{4}, 161 (2010).

\bibitem{lei1998spectroscopic}
G.~Lei, J.~E. Anderson, M.~I. Buchwald, B.~C. Edwards, R.~I. Epstein, M.~T.
  Murtagh, and G.~Sigel, \enquote{Spectroscopic evaluation of {Y}b$^{3+}$-doped
  glasses for optical refrigeration,} {{IEEE journal of
  quantum electronics}} \textbf{34}, 1839--1845 (1998).

\bibitem{melgaard2010spectroscopy}
S.~Melgaard, D.~Seletskiy, M.~Sheik-Bahae, S.~Bigotta, A.~Di~Lieto, M.~Tonelli,
  and R.~Epstein, \enquote{Spectroscopy of {Y}b-doped {YLF} crystals for laser
  cooling,} in \emph{Laser Refrigeration of Solids III,} , vol. 7614
  (International Society for Optics and Photonics, 2010), p. 761407.

\bibitem{ruan2006enhanced}
X.~Ruan and M.~Kaviany, \enquote{Enhanced laser cooling of rare-earth-ion-doped
  nanocrystalline powders,} {{Physical Review B}}
  \textbf{73}, 155422 (2006).

\bibitem{hehlen2007model}
M.~P. Hehlen, R.~I. Epstein, and H.~Inoue, \enquote{Model of laser cooling in
  the {Y}b$^{3+}$-doped fluorozirconate glass {ZBLAN},}
  {{Physical Review B}} \textbf{75}, 144302 (2007).

\bibitem{digonnet2001rare}
M.~J. Digonnet, \emph{Rare-earth-doped fiber lasers and amplifiers, revised and
  expanded} (CRC press, 2001).

\bibitem{hoyt2003advances}
C.~Hoyt, M.~Hasselbeck, M.~Sheik-Bahae, R.~Epstein, S.~Greenfield, J.~Thiede,
  J.~Distel, and J.~Valencia, \enquote{Advances in laser cooling of
  {T}hulium-doped glass,} {{JOSA B}} \textbf{20},
  1066--1074 (2003).

\bibitem{barua2008influences}
P.~Barua, E.~Sekiya, K.~Saito, and A.~Ikushima, \enquote{Influences of
  {Y}b$^{3+}$ ion concentration on the spectroscopic properties of silica
  glass,} {{Journal of Non-Crystalline Solids}}
  \textbf{354}, 4760--4764 (2008).

\bibitem{van1983nonradiative}
J.~Van~Dijk and M.~Schuurmans, \enquote{On the nonradiative and radiative decay
  rates and a modified exponential energy gap law for 4 f--4 f transitions in
  rare-earth ions,} {{The Journal of Chemical Physics}}
  \textbf{78}, 5317--5323 (1983).

\bibitem{auzel2003radiation}
F.~Auzel, G.~Baldacchini, L.~Laversenne, and G.~Boulon, \enquote{Radiation
  trapping and self-quenching analysis in {Y}b$^{3+}$, {E}r$^{3+}$, and
  {H}o$^{3+}$ doped {Y}$_{2}${O}$_{3}$,} {{Optical
  Materials}} \textbf{24}, 103--109 (2003).

\bibitem{boulon2008so}
G.~Boulon, \enquote{Why so deep research on {Y}b$^{3+}$-doped optical inorganic
  materials?} {{Journal of Alloys and Compounds}}
  \textbf{451}, 1--11 (2008).

\bibitem{nguyen2012all}
D.~T. Nguyen, J.~Zong, D.~Rhonehouse, A.~Miller, Z.~Yao, G.~Hardesty, N.~Kwong,
  R.~Binder, and A.~Chavez-Pirson, \enquote{All fiber approach to solid-state
  laser cooling,} in \emph{Laser Refrigeration of Solids V,} , vol. 8275
  (International Society for Optics and Photonics, 2012), p. 827506.

\bibitem{jetschke2008efficient}
S.~Jetschke, S.~Unger, A.~Schwuchow, M.~Leich, and J.~Kirchhof,
  \enquote{Efficient {Y}b laser fibers with low photodarkening by optimization
  of the core composition,} {{Optics Express}}
  \textbf{16}, 15540--15545 (2008).

\bibitem{leich2011highly}
M.~Leich, F.~Just, A.~Langner, M.~Such, G.~Sch{\"o}tz, T.~Eschrich, and
  S.~Grimm, \enquote{Highly efficient {Y}b-doped silica fibers prepared by
  powder sinter technology,} {{Optics letters}}
  \textbf{36}, 1557--1559 (2011).

\bibitem{Sidharthan:18}
R.~Sidharthan, S.~H. Lim, K.~J. Lim, D.~Ho, C.~H. Tse, J.~Ji, H.~Li, Y.~M.
  Seng, S.~L. Chua, and S.~Yoo, \enquote{Fabrication of low loss low-{NA}
  highly {Y}b-doped {A}luminophosphosilicate fiber for high power fiber
  lasers,} in \emph{Conference on Lasers and Electro-Optics,}  (Optical Society
  of America, 2018), p. JTh2A.129.


\bibitem{faure2007improvement}
B.~Faure, W.~Blanc, B.~Dussardier, and G.~Monnom, \enquote{Improvement of the
  {T}m$^{3+}$: $^{3}${H}$_{4}$ level lifetime in silica optical fibers by
  lowering the local phonon energy,} {{Journal of
  Non-Crystalline Solids}} \textbf{353}, 2767--2773 (2007).

\bibitem{thomas2000towards}
G.~A. Thomas, B.~I. Shraiman, P.~F. Glodis, and M.~J. Stephen, \enquote{Towards
  the clarity limit in optical fibre,} {{Nature}}
  \textbf{404}, 262 (2000).

\bibitem{jeong2004ytterbium}
Y.~Jeong, J.~K. Sahu, D.~N. Payne, and J.~Nilsson, \enquote{Ytterbium-doped
  large-core fiber laser with 1.36 kw continuous-wave output power,}
  {{Opt. Express}} \textbf{12}, 6088--6092 (2004).

\bibitem{fan1989end}
T.~Fan and M.~Kokta, \enquote{End-pumped {N}d:{L}a{F}$^{3+}$ and
  {N}d:{L}a{M}g{A}l$_{11}${O}$_{19}$ lasers,} {{IEEE
  journal of quantum electronics}} \textbf{25}, 1845--1849 (1989).

\bibitem{mccumber1964theory}
D.~E. McCumber, \enquote{Theory of phonon-terminated optical masers,}
  {{Physical review}} \textbf{134}, A299 (1964).

\bibitem{newell2007temperature}
T.~Newell, P.~Peterson, A.~Gavrielides, and M.~Sharma, \enquote{Temperature
  effects on the emission properties of {Y}b-doped optical fibers,}
  {{Optics communications}} \textbf{273}, 256--259 (2007).

\bibitem{aull1982vibronic}
B.~Aull and H.~Jenssen, \enquote{Vibronic interactions in {N}d:{YAG} resulting
  in nonreciprocity of absorption and stimulated emission cross sections,}
  {{IEEE Journal of Quantum Electronics}} \textbf{18},
  925--930 (1982).

\bibitem{Mobini:18}
E.~Mobini, M.~Peysokhan, B.~Abaie, and A.~Mafi, \enquote{Investigation of solid
  state laser cooling in ytterbium-doped silica fibers,} in \emph{Conference on
  Lasers and Electro-Optics,}  (Optical Society of America, 2018), p. FF3E.4.

\bibitem{zhang1993thermal}
Z.~Zhang, K.~Grattan, and A.~Palmer, {{Journal of applied
  physics}} \textbf{73}, 3493 (1993).


\bibitem{arai1986aluminum}
K.~Arai, H.~Namikawa, K.~Kumata, T.~Honda, Y.~Ishii, and T.~Handa,
  \enquote{Aluminum or phosphorus co-doping effects on the fluorescence and
  structural properties of neodymium-doped silica glass,}
  {{Journal of Applied Physics}} \textbf{59}, 3430--3436
  (1986).

\bibitem{laegsgaard2002dissolution}
J.~L{\ae}gsgaard, \enquote{Dissolution of rare-earth clusters in {S}i{O}$_{2}$
  by {A}l codoping: a microscopic model,} {{Physical
  Review B}} \textbf{65}, 174114 (2002).

\end{thebibliography}
\end{document}